\documentclass[aps,prb,twocolumn,floats,final,superscriptaddress,citeautoscript,longbibliography]{revtex4-2}
\usepackage{amsmath,amsthm,verbatim,amssymb,amsfonts,amscd,graphicx}
\usepackage[colorlinks=true,citecolor=blue,linkcolor=magenta,urlcolor=magenta]{hyperref}
\usepackage{multirow}
\usepackage{color}
\usepackage{makecell}

\newcommand{\avg}[1]{\left< #1 \right>}
\usepackage{soul}

\usepackage{bm}

\newcommand{\Ry}{R\'enyi }

\newcommand{\tr}{\text{tr}}
\renewcommand{\vec}[1]{\boldsymbol{\mathbf{#1}}}

\newcommand{\be}{\begin{equation}}
\newcommand{\ee}{\end{equation}}
\newcommand{\bea}{\begin{equation}\begin{aligned}}
\newcommand{\eea}{\end{aligned}\end{equation}}

\begin{document}

\title{Information Scrambling in Free Fermion Systems with a Sole Interaction}
\author{Qucheng Gao}
\email{gaoqc@bc.edu}
\affiliation{Department of Physics, Boston College, Chestnut Hill, MA 02467, USA}
\author{Pengfei Zhang}
\affiliation{Department of Physics, Fudan University, Shanghai, 200438, China}
\affiliation{Shanghai Qi Zhi Institute, AI Tower, Xuhui District, Shanghai 200232, China}
\author{Xiao Chen}
%\email{chenaad@bc.edu}
\affiliation{Department of Physics, Boston College, Chestnut Hill, MA 02467, USA}
\begin{abstract}

     It is well established that the presence of single impurity can have a substantial impact on the transport properties of quantum many-body systems at low temperature. In this work, we investigate a close analog of this problem from the perspective of quantum information dynamics. We construct Brownian circuits and Clifford circuits consisting of a free fermion hopping term and a sole interaction. In both circuits, our findings reveal the emergence of operator scrambling. Notably, the growth of the operator can be mapped to the symmetric exclusion process in the presence of a source term localized at a single point. We demonstrate that in the one-dimensional system, both the operator and entanglement exhibit diffusive scaling. Conversely, in scenarios characterized by all-to-all hopping, the operator's size undergoes exponential growth, while the entanglement exhibits a linear increase over time.

\end{abstract}

\maketitle

\tableofcontents

\section{Introduction}
The past few years have witnessed remarkable advancements in the realm of many-body quantum information dynamics, spanning both the integrable free fermion systems~\cite{calabrese2005evolution,fagotti2008evolution,alba2017entanglement,alba2018entanglement} and the nonintegrable interacting systems~\cite{lauchli2008spreading,kim2013ballistic,zhang2015thermalization,ho2017entanglement,nahum2017quantum,nahum2018operator,von2018operator,jonay2018coarse}. In the free fermion systems, the spreading of the information hinges upon the propagation of quasiparticle pairs~\cite{calabrese2005evolution,calabrese2016quantum}. Conversely, in an interacting system, the dynamics are governed by information scrambling~\cite{hayden2007black,sekino2008fast,shenker2014black,shenker2014multiple,shenker2015stringy,maldacena2016bound,hosur2016chaos,roberts2018operator,chen2018operator,xu2022scrambling}. A seemingly simple operator, when subjected to time evolution, can evolve into an exceedingly complex operator---this phenomenon is famously referred to as the quantum butterfly effect~\cite{shenker2014black}. Motivated by these developments, in this paper, we explore the quantum information dynamics in a one-dimensional free fermion system with only a sole interaction.

Similar problem has been investigated in the conventional condensed matter physics, which is known as the Kondo problem, where itinerant fermions interact with a sole magnetic impurity~\cite{kondo1964resistance,giamarchi2003quantum}. Even when the interaction initially begins as weak, the metallic phase can experience significant modifications if the interaction becomes relevant under the renormalization group flow, resulting in an anomalous increase in resistivity as the temperature decreases. The insights gained from studying the Kondo problem have far-reaching implications in various areas, including heavy-fermion systems, quantum dots, and topological materials.

To understand the information dynamics in a free fermion system with only a sole interaction, we consider the Heisenberg evolution of a simple operator. For the free fermion system, the operator remains simple under time evolution. However, for an interacting system,  as time progresses, the operator can transform into a highly entangled and complicated non-local operator. This so-called  operator scrambling dynamics can be quantitatively measured by the out-of-time-ordered (OTO) commutator~\cite{larkin1969quasiclassical,kitaev2015simple},
\bea
    C(t) = \avg{|[O(t),Y]|^2},
\eea
where $O(t) = U^{\dagger}(t)OU(t)$ is the Heisenberg operator and  $\avg{\dots}$ represents the thermal average at infinite temperature. The commutator should be replaced by an anticommutator if both $O$ and $Y$ are fermionic operators. In some large-$N$ systems with all-to-all interactions, the operator shows exponential growth at early time \cite{maldacena2016bound}. Concrete examples include the Sachdev-Ye-Kitaev (SYK) model \cite{sachdev1993gapless,kitaev2015simple,PhysRevD.94.106002} and its Brownian version \cite{saad2018semiclassical}. For qubit systems evolved under local unitary dynamics, the operator size grows linearly in time, with the front diffusively broadening~\cite{von2018operator,nahum2018operator,jonay2018coarse}.

In this work, we concentrate on scenarios where the interaction randomly fluctuates in the temporal dimension. Our findings reveal that in such instance, the quasi-particle picture~\cite{calabrese2005evolution,calabrese2016quantum} in the free fermion system breaks down. The solitary interaction term can induce information scrambling, albeit at a significantly slower pace. We obtain these results based on the two solvable models: (i) We construct Brownian circuits that incorporate time-dependent randomness.
By extending techniques developed in Ref.~\cite{lashkari2013towards,zhou2019operator,chen2019quantum,zhou2020operator,xu2019locality}, we derive a generic master equation for the operator dynamics and simulate an associated classical stochastic process. (ii) We build Clifford circuits, in which both operator spreading and entanglement growth can be efficiently simulated by monitoring the evolution of stabilizers~\cite{nielsen2010quantum,gottesman1998heisenberg,aaronson2004improved}. 
The results show that both the operator and entanglement exhibit diffusive scaling.
Furthermore, we consider both Brownian and Clifford circuits with all-to-all hoppings. We show that the operator experiences exponential growth at early times, while the entanglement entropy grows linearly. We anticipate that these findings describe generic features of systems with a single interaction, accompanied by stochastic fluctuations.

\section{Brownian circuits in 1D}
We first construct solvable Brownian circuits in one dimension (1D)~\cite{lashkari2013towards,zhou2019operator,chen2019quantum,zhou2020operator}. We focus on Majorana fermions with nearest-neighbor hopping and a sole four-body interaction term on the first four sites. The Hamiltonian reads 
\be\label{eq:meq1}
dH(t) = \sum_{i} dW_{i,i+1}(i\gamma_i\gamma_{i+1}) 
+ dW_{1234}(\gamma_1\gamma_2\gamma_3\gamma_4),
\ee
where $\gamma_i$ are Majorana fermion operators satisfying $\{\gamma_i,\gamma_j\} = 2\delta_{ij}$. Independent Brownian variables $dW_{i,i+1}$ and $dW_{1234}$ satisfy the Wiener process, with 
\bea\label{eq:meq2}
    \overline{dW_{i,i+1}dW_{i',i'+1}}&= Adt \delta_{ii'},\\
    \overline{dW_{1234}dW_{1234}}&= Bdt.
\eea
In a short interval $dt$, the unitary evolution is given by $dU=e^{-i dH}$. This setup is similar to Ref.~\cite{andreanov2023dyson}, in which exact diagonalization was used to study level repulsion, characterizing the many-body chaos transition of a similar model on graphs. 
Our main interest is the dynamics of simple operators. Following Refs.~\cite{zhou2019operator,chen2019quantum,zhou2020operator}, we expand $O(t)$ in a complete orthonormal basis of Hermitian operators $\{B_\mu\}=\{i^{q(q-1)/2}\gamma_{i_1}\gamma_{i_2}...\gamma_{i_q}\}$ ($1\leq i_1< i_2<...<i_q\leq L$, where $L$ is the number of sites):
\be
    O(t) = \sum_{\mu } \alpha_{\mu}(t)  B_\mu.
\ee
Here, the sum of $\mu$ is over all $2^L$ elements in the basis. $\alpha_{\mu}(t)$ can be viewed as a wavefunction in the operator space and the expression of $\alpha_{\mu}(t)$ is
\be
    \alpha_{\mu}(t) = \frac{1}{\tr( B_\mu B_\mu ) }\tr( B_{\mu} O(t) ).
\ee
With $\langle O^{\dagger}O\rangle =1$, the coefficients $|\alpha_{\mu}(t)|^2$ satisfy the normalization condition $\sum_\mu |\alpha_{\mu}(t)|^2=1$. We can assign a height for each basis operator as follows:
the $i$th component $h_i$ for operator $B_\mu$ is $0$ if $B_{\mu}$ 
is identity on site $i$ and $1$ if it is the $\gamma_i$ operator. Therefore, $h_i$ forms an $L$-component vector $\vec h \in \{0,1\}^L$ and
the site-resolved height distribution $f({\vec h},t)$ can be defined as
\bea\label{eq::defoff}
    f({\vec h},t) =|\alpha_\mu(t)|^2 \Big| _{\text{height}(B_\mu)=\vec{h}},
\eea
whose change in an infinitesimal amount of time $dt$ is given by
\bea\label{eq::added1}
    d f( B_\mu, t ) &= 2 \overline{\alpha_\mu(t) d \alpha_\mu(t) } + \overline{ d \alpha_\mu(t) d \alpha_\mu(t) }.
\eea
We can further introduce the  operator size by $\langle h(t)\rangle \equiv \sum_{\vec h} h f({\vec h},t) $ 
 with $h$ counting the value $1$ in ${\vec h}$,  which is equal to the summation over OTO correlators~\cite{roberts2018operator,zhang2023operator,qi2019quantum}
\begin{equation}
    \langle h(t)\rangle =\frac{1}{4}\sum_i\avg{|\{O(t),\gamma_i\}|^2}.
\end{equation}
For example, for operator $O = \frac{1}{\sqrt{2}}(\gamma_1+i\gamma_1\gamma_2)$, $f({\vec h}=\{1,0,0,\dots\}) = f({\vec h}=\{1,1,0,\dots\}) = 1/2$ and the operator size is $\langle h\rangle = 1/2(1+2)=3/2$.

For an initial Majorana fermion operator $O=\gamma_j$, the height distribution is a delta function at ${\vec h}=(0,0...,h_j=1,0,...)$ and $h(0)=1$. Under chaotic evolution, the height distribution approaches a uniform distribution in the  operator space in the long-time limit, with the restriction of fermion parity. The change of the height distribution at time $t$ is determined by using the Heisenberg equation
\bea\label{eq:a1}
    dO(t) = e^{idH(t)} O(t) e^{-idH(t)} - O(t).
\eea
In the context of Brownian circuits, a significant simplification arises from the dephasing effect induced by the time-dependent randomness, and we can expand Eq.~\eqref{eq:a1} and keep terms to the second order,
\bea\label{eq:a2}
    dO(t) = [ i dH(t), O (t) ]  + \frac{1}{2} [ i dH(t), [ i dH(t), O (t) ] ].
\eea
With the expression of $dO(t)$, the time evolution of the expansion coefficient $\alpha_{\mu}(t)$ is
\bea\label{eq:a3}
    d \alpha_{\mu} (t) = \frac{1}{\tr( B_\mu^2) }  \tr( B_{\mu} d O(t) )
\eea
Plugging this into the change of the height distribution~\eqref{eq::added1} and using Eq.~\eqref{eq:meq1} leads to a closed form for the Markovian dynamics of $f({\vec h},t)$ described by a master equation. Leaving the details to the Appendix~\cite{sup}, the result reads
\bea\label{eq:meqint}
    \frac{\partial f( \vec{h}, t ) }{\partial t}  
    &=4A\sum_{i}\delta_{h_i \oplus h_{i+1},1} \\
    & \quad \Big( 
    f( \vec{h} \oplus \vec{e}_i \oplus \vec{e}_{i+1} , t ) -f( \vec{h}, t )\Big)\\
    &+4B~\delta_{h_1 \oplus h_2 \oplus h_3\oplus h_4,1}\\ 
    &\quad \Big(     f( \vec{h} \oplus \vec{e}_1 \oplus \vec{e}_{2} \oplus \vec{e}_{3} 
    \oplus \vec{e}_{4} , t )-f(\vec{h},t) \Big),
\eea
where $\vec{e}_i$ represents an $L$-component vector that takes the value $1$ on site $i$ and $0$ at all other sites. The sum $\oplus$ is taken modulo 2.  

The first term on the right-hand side of Eq.~\eqref{eq:meqint} represents the hopping of fermions. Any $\gamma_i$ operator in the original configuration $\vec{h}$ can move one site to the left or right with a rate of $4A$, provided that the target site is unoccupied. Since there is no change in the number of Majorana operators, this term preserves the operator size. Indeed, this corresponds to the symmetric exclusion process (SEP) ~\cite{liggett1985interacting,spitzer1991interaction} and can be reduced to an unbiased random walk problem when there is only a single fermion operator~\cite{sup}. The second term is the contribution from the interaction term and only has a nontrivial operation on bases that contain one or three Majorana fermions on the first four sites. This constraint is described by the Kronecker delta function. As an example, $\vec{h} = \{1,0,0,0,0, \dots\}$ can transform to $\vec{h}_{\text{new}} = \{0,1,1,1,0, \dots\}$ under the operation of the second term. This results in an increase in the operator size by $2$.

\begin{figure}
    \centering
    \includegraphics[width=0.98\columnwidth]{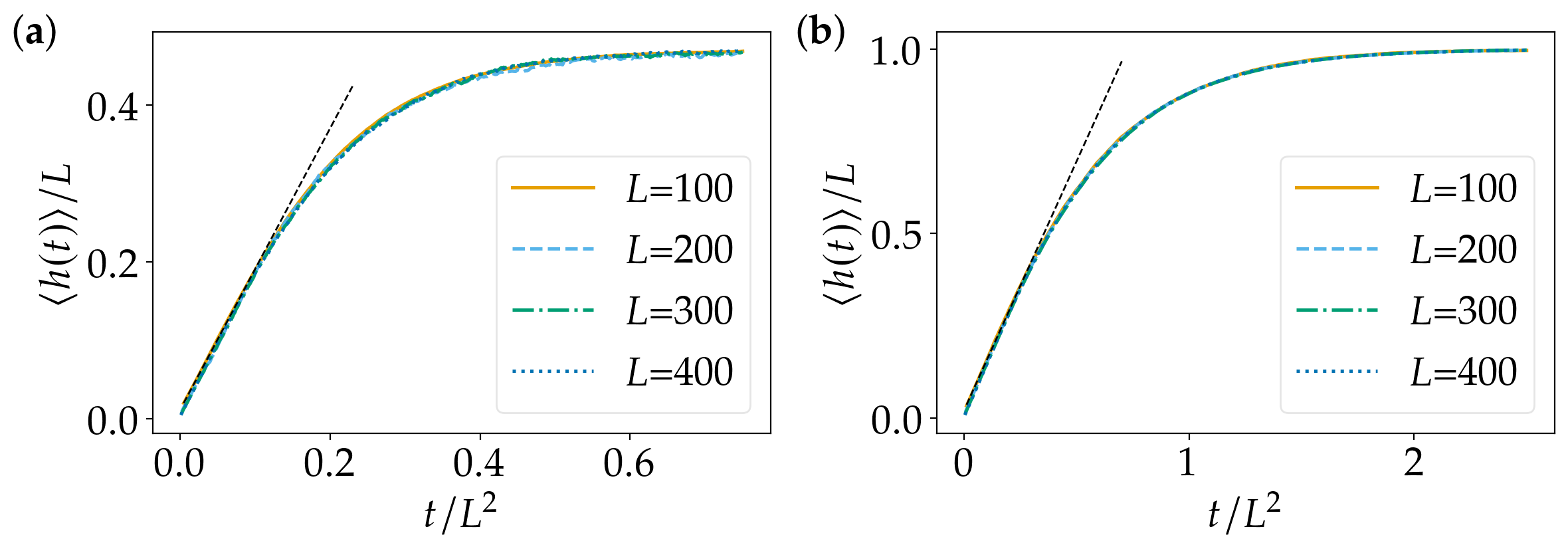}
    \caption{Operator size growth in models with spatial locality. (a) Operator size growth of 1D Brownian circuits with  $A=B=1/4$. The results are obtained by classical simulation. 
    (b) Operator size growth of 1D Clifford circuits. 
    The black dashed lines in both figures indicate linear behavior. The data collapse demonstrates a clear diffusive scaling in both models. In both models, the operator's initial location is randomly selected from a uniform distribution across the lattice.
    Note that in Clifford circuits, $L$ is the number of qubits and there are $2L$ Majorana fermions.
    }
    \label{int_fig}	
\end{figure}

$f( \vec{h}, t )$ contains exponentially many components, which makes a direct analytical or numerical study of Eq.~\eqref{eq:meqint} for large system size impractical. Therefore, we first construct a classical stochastic particle dynamics described by Eq.~\eqref{eq:meqint} and then perform a Monte Carlo simulation to understand operator growth. (In the Appendix~\cite{sup}, we compare exact results with classical simulations for small systems.)
The classical model is defined on a 1D chain with $L$ empty sites and periodic boundary condition. Each site can have at most one particle. At the beginning, we put a particle on a randomly chosen site, which corresponds to the initial condition with a single Majorana operator. The dynamics of the particles is governed by the following update rule:
\begin{enumerate}
    \item If there are one or three particles on the first four sites,
    we empty the occupied sites and occupy the remaining empty sites with probability $p_B=4B \delta t$.
    \item Implement a random walk on each particle with probability $p_A=4A \delta t$.
    \item Repeat step (1) and step (2).
\end{enumerate}
This classical model can be understood as particles undergoing a SEP 
with a source at one point, which can change the number of particles. We then compute the particle number after $2N$ steps, which corresponds to the operator size at $t=N\delta t$.

The results for different system size $L$ are presented in Fig.~\ref{int_fig}(a). We observe that the particle number grows linearly at early times and approaches a nearly half-filled state in the long-time limit. In addition, we plot $\langle h(t)\rangle /L$ as a function of $t/L^2$, and find nice data collapse for different $L$ for the entire dynamics.

\begin{figure}
    \centering
    \includegraphics[width=0.98\columnwidth]{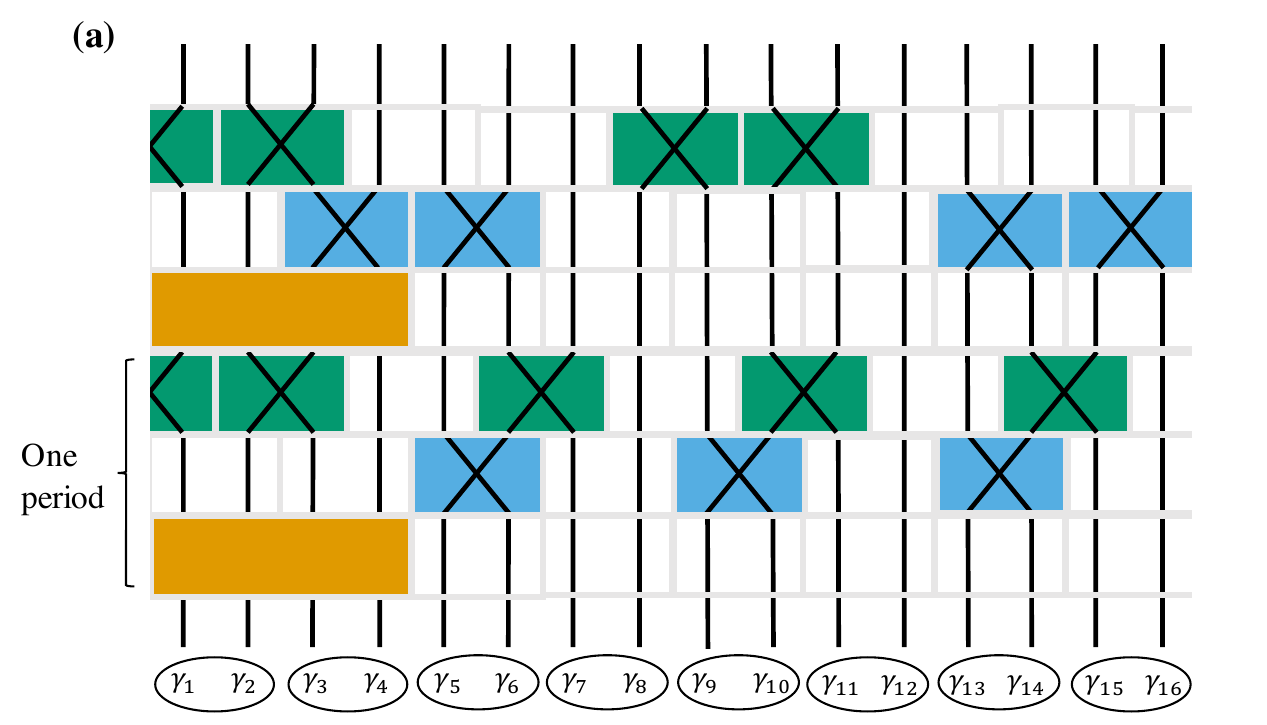}
    \includegraphics[width=0.80\columnwidth]{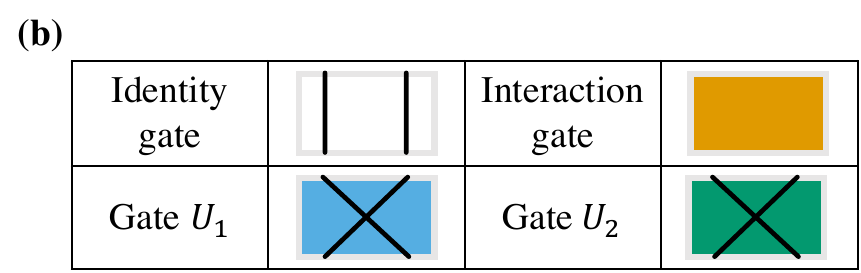}
    \caption{(a) The cartoon picture for the 1D Clifford circuits in the Majorana fermion representation. The initial state
    is stabilized by pairing up $\{(\gamma_1, \gamma_2),(\gamma_3, \gamma_4), \dots, (\gamma_{2L-1},\gamma_{2L})\}$.
    In the simulation, we take the periodic boundary condition.
    (b) The possible operations in the Clifford circuits.}
    \label{circuit}	

\end{figure}

\begin{figure}
    \centering
    \includegraphics[width=0.98\columnwidth]{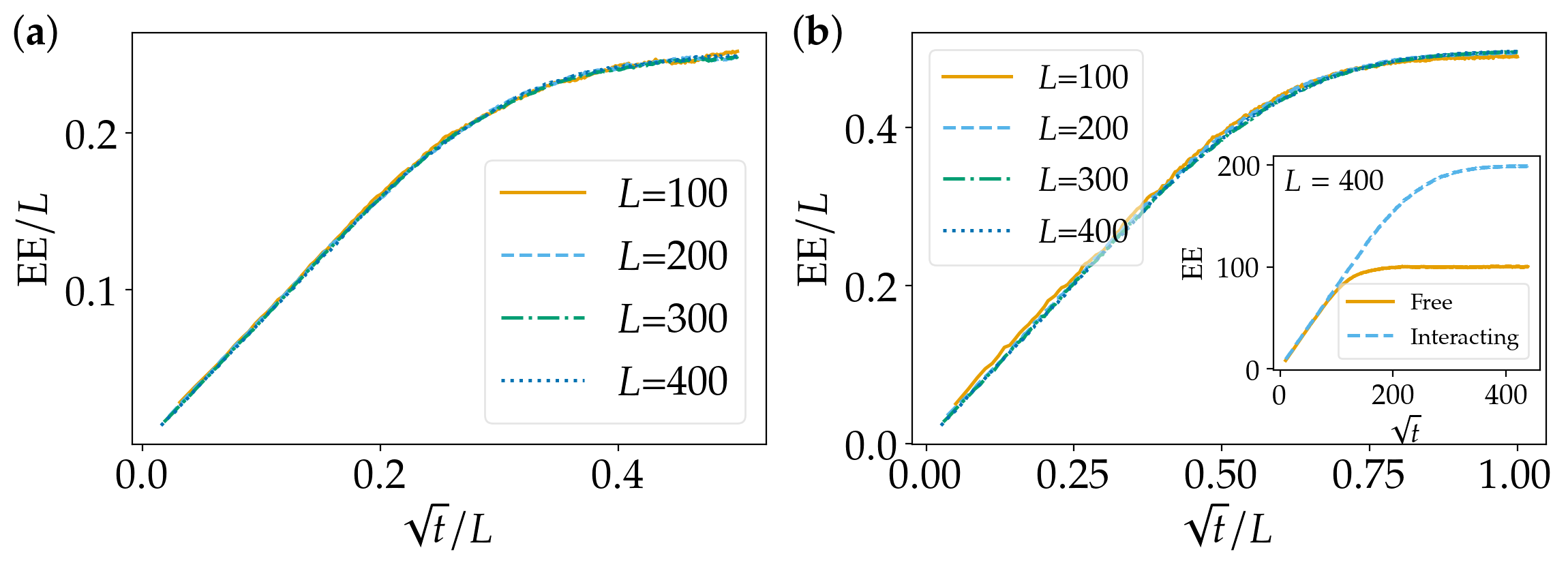}
    \caption{Entanglement growth of subsystem $A$ in 1D Clifford circuits with periodic boundary condition. We present results for (a) the free model and (b) the free model with a single random two-qubit gate. Inset: A comparison between these two models. We bipartite the system into $A$ and $\overline{A}$. The subsystem $A$ is chosen from the $(L/4+2)$-th qubit to the $(3L/4+1)$-th qubit so that the two-qubit interacting gate lies in the middle of the subsystem $\overline{A}$.
    }
    \label{free_fig1}	
\end{figure}

\section{Clifford circuits in 1D}
To explore the universality of the dynamics witnessed within the Brownian circuit, we introduce a second model based on the Clifford circuits~\cite{sang2021entanglement}. Like the Brownian circuits, this model also describes Majorana fermions with the interaction taking place exclusively among the first four sites. Unlike a continuous Brownian Hamiltonian evolution, the system progresses through a sequence of Clifford gates, facilitating a stabilizer representation for the quantum state. In a system comprising $L$ qubits, this necessitates $L$ distinct Pauli string operators, wherein the quantum state manifests as an eigenstate with an eigenvalue of $+1$. As a result, the Clifford dynamics can be translated into the classical dynamics of stabilizers, making it amenable to efficient simulation on classical computers \cite{gottesman1998heisenberg,aaronson2004improved}.

 We consider a setup with $2L$ Majorana fermions, which corresponds to an $L$-qubit system. The Pauli operator is related to the Majorana fermion representation by the Jordan-Wigner transformation,
\bea
    X_i = i\gamma_{2i-1}\gamma_{2i},\ \ \ \ \ \ 
    Y_i = \prod_{j<i}X_j\gamma_{2i}.
\eea
According to this transformation, a Pauli string can be mapped to a Majorana string up to a phase factor.
The initial state is stabilized by $\{i\gamma_1\gamma_2,i\gamma_3\gamma_4, \dots,i\gamma_{2L-1}\gamma_{2L}\}$, which corresponds to pairing up $\gamma_{2i-1}$ and $\gamma_{2i}$. In this quantum dynamics, each time period is composed of three steps (also see Fig.~\ref{circuit}):
\begin{enumerate}
    \item Apply a random two-qubit Clifford gate on the first two qubits, which corresponds to an  interaction term in the Majorana fermion representation.

    \item Perform a swap $U_1 =  e^{i\frac{\pi}{4} (i\gamma_{2i-1}\gamma_{2i}) }$ between Majorana fermion modes $\gamma_{2i-1}$ and $\gamma_{2i}$ with probability $p$. This corresponds to a Clifford gate.

    \item Perform a swap $U_2 =  e^{i\frac{\pi}{4} (i\gamma_{2i}\gamma_{2i+1}) }$ between Majorana fermion modes $\gamma_{2i}$ and $\gamma_{2i+1}$ with probability $p$. This is also a Clifford gate. Steps (2) and (3) together describe a free Majorana fermion dynamics. 
\end{enumerate}

We focus on the growth of the stabilizer operator, computed in the Majorana fermion representation.
The results are depicted in Fig.~\ref{int_fig}(b). In the early stages, the operator size shows a linear growth, with the growth rate proportional to $1/L$. Furthermore, $\langle h(t)\rangle/L$ is a scaling function $f(t/L^2)$ for the entire dynamics. Both the early-time linear growth and the diffusive scaling are consistent with the results observed in the classical model extracted from the Brownian circuit. In both models, the operator dynamics can be understood in terms of SEP in the presence of a source term at the boundary.

We are also interested in the quantum entanglement growth in this Clifford dynamics.
The \Ry entanglement entropy (EE) of a subsystem $A$ is defined as
\be
    S_A^{(n)} = \frac{1}{1-n}\log\tr\rho_A^n,
\ee
where $\rho_A$ is the reduced density matrix
of the subsystem $A$. In this paper, we use the logarithm with base $2$. For the stabilizer state, EE is independent of the \Ry index and therefore we can ignore the superscript $n$. Unlike the operator dynamics, which only grows in the presence of the interaction, the results in Fig.~\ref{free_fig1} exhibit $\sqrt{t}$ growth  for both interacting and noninteracting cases [evolved without step (1)]. However, the entanglement structures of these two cases are quite different. In interacting systems, the half-system EE eventually saturates to approximately half of the system size, constituting the maximum possible EE for the half system. In noninteracting systems, it only saturates to about one-fourth of the system size. This can be explicitly understood by counting the number of fermion arcs crossing the entanglement cut. Each arc of a fermion-pair stabilizer $(\gamma_i,\gamma_j)$ has a $1/2$ probability of connecting region $A$ to its exterior, contributing $1/2$ to the EE. Consequently, roughly half of the $L$ stabilizers will collectively contribute $L/2 \cdot 1/2 = L/4$ for EE. For completeness, we also investigate a Floquet free fermion Clifford circuit with a fluctuating sole interaction in the Appendix~\cite{sup}.

\begin{figure}
    \centering
    \includegraphics[width=0.98\columnwidth]{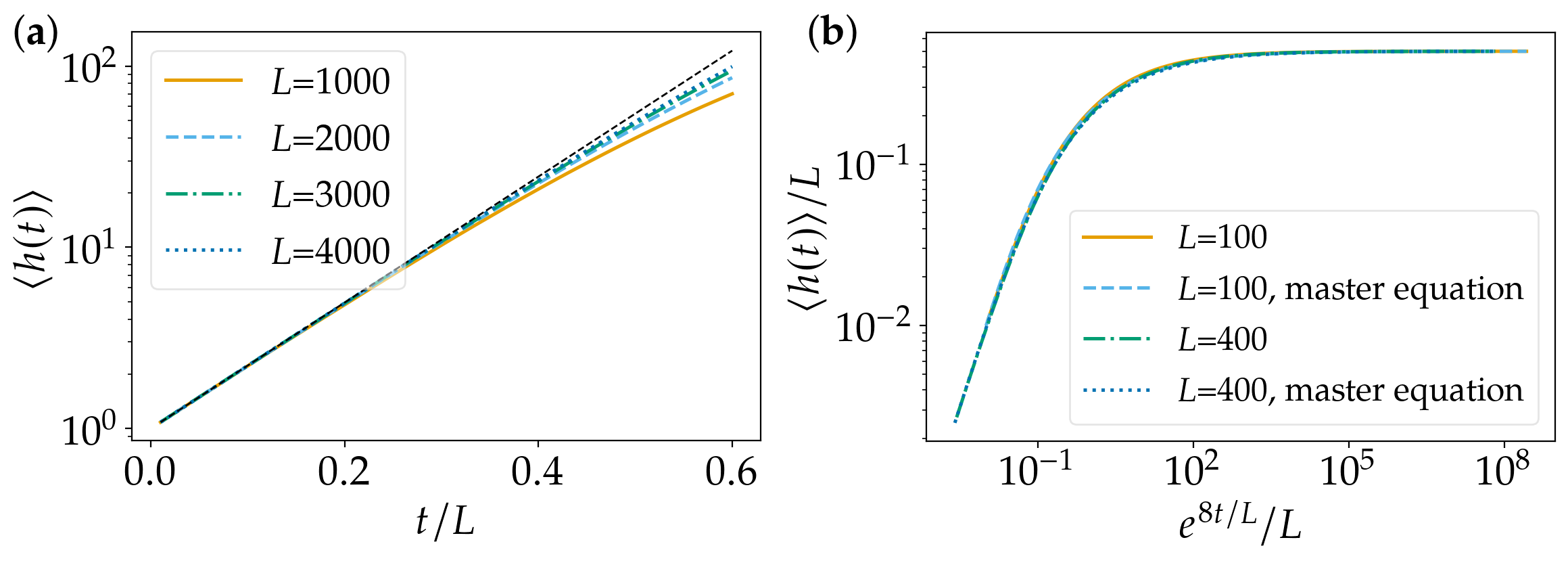}
    \caption{Operator size growth of nonlocal 1D Brownian circuits with $B=1/4$. (a) The results are obtained by classical simulation,
    which show that the operator size grows exponentially as a function of $t/L$ at early times. The black dashed line indicates $\avg{h} = e^{8 t/L}$.
    (b) The comparison between the solution of the master equation~\eqref{eq:bmeq_nl} and the results from the classical particle simulation. 
    }
    \label{bnl_fig}	
\end{figure}

\begin{figure}
    \centering
    \includegraphics[width=0.98\columnwidth]{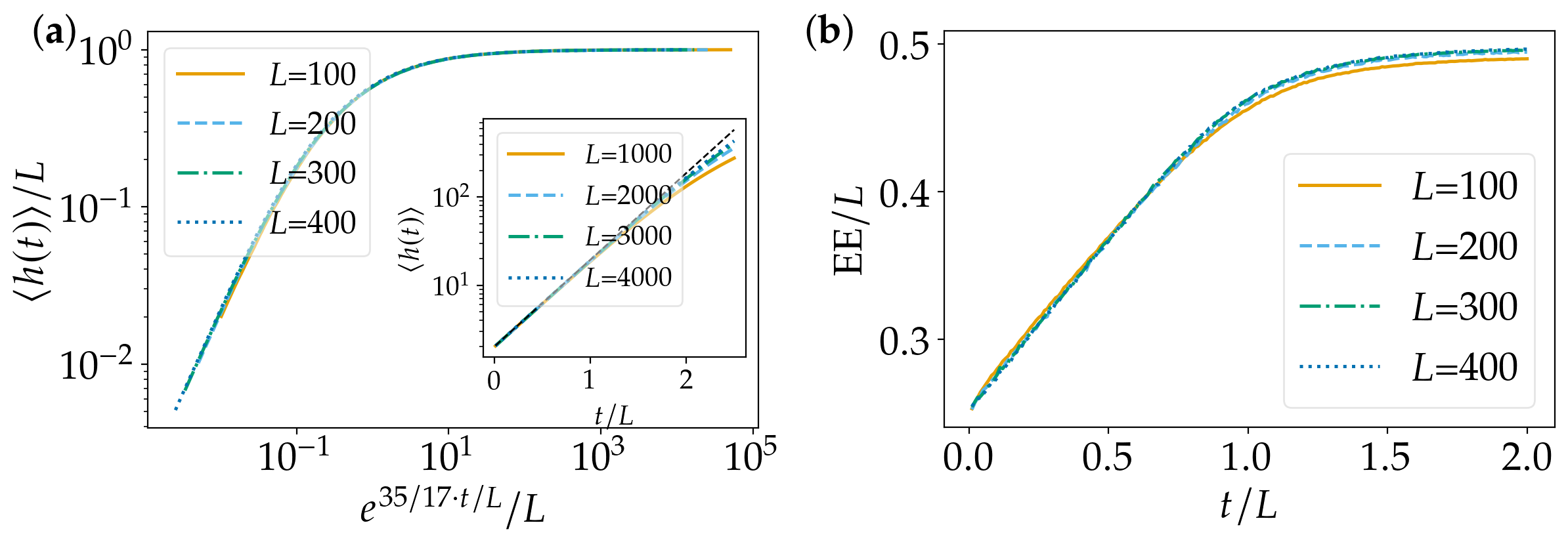}
    \caption{Non-local model in 1D Clifford circuits. (a) Operator size growth.
    The inset shows that the operator size grows exponentially as a function of $t/L$ at early times;
    (b) The data collapse of entanglement growth of subsystem $A$ for different $L$. The size of subsystem $A$ is half of the whole system.
    }
    \label{nl_fig}	
\end{figure}

\section{Nonlocal model}
Here we turn to models with all-to-all hopping. In the Brownian circuit, the Hamiltonian is now
\be\label{eq:meq3}
dH(t) = \sum_{i<j} dW_{i,j}(i\gamma_i\gamma_{j}) 
+ dW_{1234}(\gamma_1\gamma_2\gamma_3\gamma_4),
\ee
where $dW_{ij}$ and $dW_{1234}$ satisfy the Wiener process,
\bea\label{eq:meq4}
    \overline{dW_{ij}dW_{i'j'}}&= Adt \delta_{ii'}\delta_{jj'},\\
    \overline{dW_{1234}dW_{1234}}&= Bdt.
\eea
 For this model, the operator dynamics can once again be described by the classical master equation for the height distribution $f(\vec{h},t)$. This master equation is similar to Eq.~\eqref{eq:meqint} with the details given in the Appendix~\cite{sup}. It describes the classical nonlocal random walker in the presence of a source term. Therefore the growth of the particle/operator dynamics is governed by the following update rule:
\begin{enumerate}
    \item If there are one or three particles on the first four sites,
    empty the occupied sites and occupy the remaining empty sites with probability $p_B=4B \delta t$.
    \item Implement a reshuffle on all sites. Specifically, we randomly permutate the $L$ sites.
    \item Repeat step (1) and step (2).
\end{enumerate}
 Here, the random permutation among all sites in step (2) arises from the presence of a finite constant $A$ in the large $L$ limit. This leads to a classical dynamics that is independent of $A$. For this dynamics, the particle number at $2N$ steps corresponds to the operator size at $t=N\delta t$.

The results for different system size $L$ are presented in Fig.~\ref{bnl_fig}(a). We observe that the particle number grows exponentially at early times with the growth rate $\sim 1/L$. This can be understood as follows: 
Following this nonlocal random walk process, the probability that one site is occupied by a particle is $\avg{h}/L$. At early times, when $\avg{h}/L\ll 1$, the probability of only one site among the first four sites being occupied far outweighs the probabilities associated with other nonzero particle configurations on those initial four sites. The interaction gate can increase the size of such size-one operator by a constant $2$. Since there are four such configurations, we can obtain
\bea\label{eq:meq_et}
    \frac{d\avg{h}}{dt} = 4\cdot 2 \avg{h}/L,
\eea
with $B=1/4$.
The solution of the above equation is proportional to an exponential function $\avg{h} \propto e^{8 t/L}$.

Moreover, given the absence of locality in this model, we can combine $f(\vec{h},t)$ with the same operator size and define the size distribution as follows:
\bea
    f(h,t) &\equiv \sum_{ \{\vec{h} \mid \text{size of} \ \vec{h} = h\} } f(\vec{h},t)\\
    &= f(\vec{h},t)\cdot 
    \text{number of $\vec{h}$} \Big|_{\text{size of $\vec{h} = h$}},
\eea
for which we can derive a master equation,
\be\label{eq:bmeq_nl}
    \frac{d{\vec f}(t)}{dt} = A_f{\vec f}(t),
\ee
with a pentadiagonal matrix $A_f$~\cite{sup}, where ${\vec f}(t)$ is an $L$-dimensional array with elements $f(1,t),\dots,f(L,t)$. We can directly solve this master equation. In Fig.~\ref{bnl_fig}(b), we compare the solution of the master equation~\eqref{eq:bmeq_nl} with the direct simulation of the particle dynamics and find that these two methods give the same result for the entire dynamics. We also offer an analytical solution to the master equation at early times in the Appendix~\cite{sup}, which is consistent with Eq.~\eqref{eq:meq_et}.

In Clifford circuits, a similar all-to-all model is explored, with each time step including two parts:
\begin{enumerate}
    \item Apply a random two-qubit Clifford gate on the first two qubits.
    \item Apply a reshuffle on all fermions. Specifically, we randomly permutate the $2L$ fermions.
\end{enumerate}

The growth of the stabilizer operator in the Majorana fermion representation is calculated for different system size $L$ in Fig.~\ref{nl_fig}(a). The inset of Fig.~\ref{nl_fig}(a) shows that $\avg{h}$ grows exponentially as a function of $t/L$ at early times, which is again consistent with the results in the Brownian circuit. Here, the coefficient before $t/L$ is $34/15$, which is explained in the Appendix~\cite{sup}.

We also investigate the quantum entanglement growth in the Clifford dynamics and find that the EE shows a linear growth with a speed independent of the system size and eventually saturates to the maximum possible EE for the half system [see Fig.~\ref{nl_fig}(b)].
This indicates that the stabilizers initially supported on the subsystem $A$ spread over the entire system.

\begin{table*}
    \centering
    \begin{tabular}{ |c|c|c|c|}
        \hline
        Spatial locality & Model & Operator size $\avg{h(t)}$ & Entanglement entropy (EE)\\ \hline
        
        \multirow{2}{*}{Local} & Brownian circuits [Eq.~\eqref{eq:meq1}]
        & \makecell{$\avg{h(t)}/L$ is a function of $t/L^2$ \\ for the entire dynamics, \\ $\avg{h(t)} \propto t/L $ at early times [Fig.~\ref{int_fig}(a)]} &$\slash$ \\ \cline{2-4}
        & Clifford circuits (Fig.~\ref{circuit}) & \makecell{$\avg{h(t)}/L$ is a function of $t/L^2$ \\ for the entire dynamics, \\ $\avg{h(t)} \propto t/L$ at early times [Fig.~\ref{int_fig}(b)]} & \makecell{$\text{EE}/L$ is a function of $t/L^2$ \\ for the entire dynamics, \\ EE $\propto \sqrt{t}$ at early times [Fig.~\ref{free_fig1}(b)]} \\ \hline

        \multirow{2}{*}{Nonlocal} & Brownian circuits [Eq.~\eqref{eq:meq3}]
        &$\log\avg{h(t)} \propto t/L$ at early times [Fig.~\ref{bnl_fig}(a)] & $\slash$ \\ \cline{2-4}
        & Clifford circuits  & $\log\avg{h(t)}  \propto t/L$ at early times [Fig.~\ref{nl_fig}(a)] & \makecell{$\text{EE}/L$ is a function of $t/L$ \\ for the entire dynamics, \\ EE $\propto t$ at early times [Fig.~\ref{nl_fig}(b)]} \\ \hline
    \end{tabular}
    \caption{Summary of the main results in free fermion models with a sole interaction. All the models here exhibit stochastic randomness.}
    \label{tab}
\end{table*}

\section{Discussion}
In this work, we have considered one-dimensional and all-to-all free fermion models with a sole interaction. We analyze the operator dynamics within a Brownian circuit and derive the master equation for it. Our investigation reveals that the single interaction term can induce operator scrambling, albeit more slowly than an extended interaction.
In the local model, the operator size shows a diffusive scaling, while in the extended case, the operator size shows a linear behavior~\cite{nahum2018operator,von2018operator}. In the nonlocal model, the exponent for the early dynamics is suppressed by the system size compared to the extended case~\cite{zhou2019operator}.
We verify these results with large-scale simulations in the Clifford circuit.

Additionally, we investigate the entanglement dynamics within the Clifford circuits. Our findings reveal that in the one-dimensional model, the entanglement entropy exhibits diffusive growth over time, while in the model featuring all-to-all hopping, it demonstrates linear growth over time. 
We make a summary of our main results in Table~\ref{tab}.

It is worth noting that the models under investigation in this paper exhibit stochastic randomness. We anticipate that the findings presented here can be applied to other analogous random quantum dynamics with a single interaction term. Nevertheless, a crucial question emerges: Does a similar phenomenon manifest itself in models without stochastic randomness? We leave this question for future study.

\begin{acknowledgements}
We thank Hanchen Liu, Yiqiu Han and Ethan Lake for useful discussions.
We gratefully acknowledge computing resources from Research Services at Boston College 
and the assistance provided by Wei Qiu. This research is supported in part by the Google Research Scholar Program and is supported in part by the National Science Foundation under Grant No. DMR-2219735 (Q.G. and X.C.) and by the National Natural Science Foundation of China under Grant No. 12374477 (P.Z.).
\end{acknowledgements}

\bibliography{main}

\newpage
\onecolumngrid
\pagenumbering{arabic} % resets `page` counter to 1
\setcounter{equation}{0}
\setcounter{figure}{0}
\setcounter{table}{0}

\renewcommand{\thepage}{A\arabic{page}}
\renewcommand{\thesection}{A\arabic{section}}
\renewcommand{\theequation}{A\arabic{equation}}
\renewcommand{\thefigure}{A\arabic{figure}}
\renewcommand{\thetable}{A\arabic{table}}

\appendix

\section{Master equation of height distribution in Brownian circuits}
In this appendix, we give a detailed derivation of the master equation in a generic Brownian
quantum circuit.
The Hamiltonian reads
\be
dH(t) = \sum_{i<j} dW_{ij}(i\gamma_i\gamma_j) 
+\sum_{i<j<k<l} dW_{ijkl}(\gamma_i\gamma_j\gamma_k\gamma_l),
\ee
where $W_{ij}$ and $W_{ijkl}$ satisfy the Wiener process, with
\be\label{eq:wienerp}
    \overline{dW_{ij}dW_{i'j'}}= A_{ij}dt \delta_{ii'}\delta_{jj'},\quad
    \overline{dW_{ijkl}dW_{i'j'k'l'}}= B_{ijkl}dt \delta_{ii'}\delta_{jj'}\delta_{kk'}\delta_{ll'}.
\ee

First, we consider $B_{ijkl} = 0$, which corresponds to the hopping Majorana fermions without four-body interaction.
We can expand the evolution of the operator $O(t)$ to second order:
\bea
      dO(t)  &= e^{idH(t)} O(t) e^{-idH(t)} - O(t)\\
      &=  [ i dH(t), O (t) ]  + \frac{1}{2} [ i dH(t), [ i dH(t), O (t) ] ] \\
      &= i [ dH(t), O(t) ] - \frac{1}{2} \{ dH(t) dH(t), O(t) \}  + d H(t)  O(t) d H (t) \\
      &= i [ dH(t), O(t) ] - \sum_{i<j} O(t) A_{ij} dt + 
      {\sum_{i<j} \gamma_i\gamma_j O(t) \gamma_{j}\gamma_{i} A_{ij} dt}.
\eea
We choose $B_\mu$ to be a complete orthonormal basis of Hermitian operators $\{B_\mu\}=\{i^{q(q-1)/2}\gamma_{i_1}\gamma_{i_2}...\gamma_{i_q}\}$ and the expansion coefficient $\alpha_{\mu}(t)$ is
\be
 \qquad \alpha_{\mu}(t) = \frac{1}{\tr( B_\mu B_\mu ) }\tr( B_{\mu} O(t) ).
\ee
Its time evolution is given by
\bea
    d \alpha_{\mu} (t) &= \frac{1}{\tr( B_\mu^2) }  \tr( B_{\mu} d O(t) ) \\
    &= \frac{i}{\tr( B_\mu^2) }  \tr( B_{\mu} [dH(t), O(t)]   )  \\
    &- \sum_{i<j}\alpha_\mu (t) A_{ij}dt  +  \frac{1}{\tr( B_\mu^2) }\sum_{i<j}\tr\big( B_{\mu} 
     \gamma_i\gamma_j O(t) \gamma_{j}\gamma_{i}\big)A_{ij}dt, \\
\eea
where
\bea
&\frac{1}{\tr( B_\mu^2) }\sum_{i<j}\tr\big( B_{\mu} 
\gamma_i\gamma_j O(t) \gamma_{j}\gamma_{i}\big)A_{ij}dt\\
=&\frac{1}{\tr( B_\mu^2) }\sum_{i<j}\tr\big( \gamma_{j}\gamma_{i}B_{\mu} 
\gamma_i\gamma_j O(t) \big)A_{ij}dt\\
=&\sum_{i<j}q_{\mu,i,j}\alpha_\mu (t)A_{ij}dt,
\eea
where $q_{\mu,i,j}=1$ if $\gamma_i, \gamma_j\in B_{\mu}$ 
or $\gamma_i, \gamma_j\notin B_{\mu}$;
$q_{\mu,i,j}=-1$ if $\gamma_i\in B_{\mu}, \gamma_j\notin B_{\mu}$
or $\gamma_j\in B_{\mu}, \gamma_i\notin B_{\mu}$,
and we finally get
\bea
    d \alpha_{\mu} (t)& = \frac{i}{\tr( B_\mu^2) }  \tr( B_{\mu} [dH(t), O(t)]   ) 
    - \sum_{i<j}\alpha_\mu (t) A_{ij}dt  +  \sum_{i<j}q_{\mu,i,j}\alpha_\mu (t)A_{ij}dt \\
    & =\frac{i}{\tr( B_\mu^2) }  \tr( B_{\mu} [dH(t), O(t)]   )
    - 2\sum_{\{i<j\mid q_{\mu,i,j}=-1\}}\alpha_\mu (t) A_{ij}dt.
\eea

Define $f( B_\mu, t )$ to be the average probability at time $t$,
\be\label{eq:added2}
f( B_{\mu} , t )  = \overline{|\alpha_{\mu}(t)|^2} 
= \overline{\alpha^2_\mu(t)}.
\ee
The evolution is given by
\bea
  d f( B_\mu, t ) &= 2 \overline{\alpha_\mu(t) d \alpha_\mu(t) } + \overline{ d \alpha_\mu(t) d \alpha_\mu(t)  }.
\eea
We have
\bea
d f( B_\mu, t ) 
=& 2\frac{i}{\tr( B_\mu^2) }  \tr( B_{\mu} [\overline{\alpha_\mu(t)dH(t)}, O(t)])
-4\sum_{\{i<j\mid q_{\mu,i,j}=-1\}}\overline{\alpha_\mu^2 (t)} A_{ij}dt
-\frac{1}{\tr^2( B_\mu^2) }\overline{\tr^2( B_{\mu} [dH(t), O(t)])}
\\
&- 4\frac{i}{\tr( B_\mu^2) }  \tr( B_{\mu} [\overline{\alpha_\mu (t) dH(t)}, O(t)]   )\sum_{\{i<j\mid q_{\mu,i,j}=-1\}}A_{ij}dt
+4\overline{\alpha_\mu^2 (t)}(\sum_{\{i<j\mid q_{\mu,i,j}=-1\}}A_{ij})^2dt^2
\\
=& -4\sum_{\{i<j\mid q_{\mu,i,j}=-1\}}f( B_\mu, t )A_{ij}dt
-\frac{1}{\tr^2( B_\mu^2) }\overline{\tr^2( O(t) [B_{\mu},dH(t)]   )}\\
=&-4\sum_{\{i<j\mid q_{\mu,i,j}=-1\}}f( B_\mu, t )A_{ij}dt
+\sum_{\nu}\sum_{i<j}\frac{1}{\tr^2( B_\mu^2) }\tr^2(B_\nu [B_{\mu},\gamma_i\gamma_j])
f( B_\nu, t )A_{ij}dt\\
=&-4\sum_{\{i<j\mid q_{\mu,i,j}=-1\}}f( B_\mu, t )A_{ij}dt
+4\sum_{\{i<j\mid q_{\mu,i,j}=-1\},\;
\{\nu\big| |B_\mu\gamma_i\gamma_j|=|B_\nu|\}}
f( B_\nu, t )A_{ij}dt,
\eea
where, on the right of the first equality, the first and fourth terms are zero from Wiener process~\eqref{eq:wienerp}, and the fifth term is ignored when considering $dt$ to the first order. The second equality uses the definition of the average probability~\eqref{eq:added2} (height distribution of $B_\mu)$ and the cyclicity of trace,
\be
\tr( B_{\mu} [dH(t), O(t)]) = \tr( O(t) [B_{\mu},dH(t)]).
\ee

Now we can consider the operator height distribution function,
\bea
    f({\vec h},t) =|\alpha_\mu(t)|^2 \Big| _{\text{height}(B_\mu)=\vec{h}}.
\eea
which satisfies the master equation,
\bea\label{eq:meqfree}
    \frac{\partial f( \vec{h}, t ) }{\partial t}
    &=-4\sum_{i<j}\delta_{h_i \oplus h_{j},1}A_{ij}f( \vec{h}, t )
    +4\sum_{i<j}\delta_{h_i \oplus h_{j},1}A_{ij}f( \vec{h} \oplus \vec{e}_i \oplus \vec{e}_j , t )\\
    &=4\sum_{i<j}\delta_{h_i \oplus h_{j},1}A_{ij}
    \Big(f( \vec{h} \oplus \vec{e}_i \oplus \vec{e}_j , t ) -f( \vec{h}, t )
    \Big).
\eea

\subsection{Symmetric exclusion process (SEP)}
When $A_{ij} = A\delta_{j-i,1}$, Eq.~\eqref{eq:meqfree} reduces to
\bea\label{eq:meqfree2}
    \frac{\partial f( \vec{h}, t ) }{\partial t}  
    &=4A\sum_{i}\delta_{h_i \oplus h_{i+1},1}
    \Big( f( \vec{h} \oplus \vec{e}_i \oplus \vec{e}_{i+1} , t ) - f( \vec{h}, t )
    \Big).
\eea
We can consider an initial operator $\gamma_{i}$, whose corresponding height distribution
is $f( \vec{e}_i, 0 )=1$, plug it into Eq.~\eqref{eq:meqfree2}, and we get
\bea\label{eq:diffusion}
    \frac{\partial f( \vec{e}_i, 0 ) }{\partial t}
    &=4Af( \vec{e}_{i-1}, 0 ) + 4Af( \vec{e}_{i+1}, 0 ) - 8Af( \vec{e}_i, 0 )\\
    &=4A\Big(
        \big[f( \vec{e}_{i+1}, 0 )-f( \vec{e}_i, 0 )\big]
        -\big[f( \vec{e}_i, 0 )-f( \vec{e}_{i-1}, 0 )\big]
    \Big)\\
    &\approx 4A\frac{\partial^2 f( \vec{e}_i, 0 ) }{\partial x^2},
\eea
where
\be
\frac{\partial f( \vec{e}_i, 0 ) }{\partial x}\equiv \displaystyle{\lim_{\Delta x \to 0}}
\frac{f( \vec{e}_{i+\Delta x},0) - f( \vec{e}_{i-\Delta x}, 0 ) }{2\Delta x}.
\ee
This diffusion equation~\eqref{eq:diffusion} indicates that the hopping model can be 
described by a SEP model.

\subsection{Master equation for generic Brownian Hamiltonian}
When $B_{ijkl}\neq 0$, we can derive a more generic master equation including four-body interaction:
\bea\label{eq:meqint2}
    \frac{\partial f( \vec{h}, t ) }{\partial t} 
    &=4\sum_{i<j}\delta_{h_i \oplus h_{j},1}A_{ij}\big(
        f( \vec{h} \oplus \vec{e}_i \oplus \vec{e}_j, t) - f( \vec{h}, t )
    \big)\\
    &\quad +4\sum_{i<j<k<l}\delta_{h_i \oplus h_{j} \oplus h_{k} \oplus h_{l},1}B_{ijkl}\big( 
        f( \vec{h} \oplus \vec{e}_i \oplus \vec{e}_j \oplus\vec{e}_k \oplus\vec{e}_l, t )
        -f( \vec{h}, t )
    \big).
\eea

If we let $A_{ij} = A\delta_{j-i,1}$ and $B_{ijkl}=B\delta_{i1}\delta_{j2}\delta_{k3}\delta_{l4}$, we get the master equation for Majorana fermions with nearest-neighbor hopping and a sole four-body interaction,
\bea
    \frac{\partial f( \vec{h}, t ) }{\partial t} 
    &=4A\sum_{i}\delta_{h_i \oplus h_{i+1},1}\big(
        f( \vec{h} \oplus \vec{e}_i \oplus \vec{e}_{i+1}, t) - f( \vec{h}, t )
    \big)\\
    &\quad +4B\delta_{h_1 \oplus h_{2} \oplus h_{3} \oplus h_{4},1}\big( 
        f( \vec{h} \oplus \vec{e}_1 \oplus \vec{e}_2 \oplus\vec{e}_3 \oplus\vec{e}_4, t )
        -f( \vec{h}, t )
    \big).
\eea

\subsection{Master equation for nonlocal model}
When $A_{ij} = A$, $B_{ijkl}=B\delta_{i1}\delta_{j2}\delta_{k3}\delta_{l4}$, 
Eq.~\eqref{eq:meqint2} describes an all-to-all hopping model with a sole four-body interaction. Given the absence of locality in the model, the height distribution function $f( \vec{h}, t )$
now only depends on the size $h$. The master equation is now given by
\bea
    \frac{\partial f( \vec{h}, t ) }{\partial t}
    = 4B\delta_{h_1 \oplus h_{2} \oplus h_{3} \oplus h_{4},1}\big( 
        f( \vec{h} \oplus \vec{e}_1 \oplus \vec{e}_2 \oplus\vec{e}_3 \oplus\vec{e}_4, t )
        -f( \vec{h}, t )
        \big).
\eea
The master equation of the size distribution $f(h,t)$ is 
\be
    \frac{d{\vec f}(t)}{dt} = A_f{\vec f}(t),
\ee
with a pentadiagonal matrix $A_f$:
\bea
    &(A_f)_{h,h-2} = 4B\cdot p(h-2,1)\\
    &(A_f)_{h,h-1}  = 0\\
    &(A_f)_{h,h}  = -4B\cdot\big( p(h,1) + p(h,3) \big)\\
    &(A_f)_{h,h+1}  = 0\\
    &(A_f)_{h,h+2}  = 4B\cdot p(h+2,3),
\eea
where
\be
    p(h,i)\equiv {4 \choose i}{L-4 \choose h-i}\bigg/{L \choose h}.
\ee

At early times, or large $L$ limit, the master equation simplifies to
\bea
    \frac{df(h,t)}{dt} = \frac{16B}{L}[(h-2)f(h-2,t)-hf(h,t) ].
\eea
The coefficient of $f(h,t)$ is the rate of height increase, which is proportional to the height itself, indicating an initial exponential growth. A similar equation was proposed for the height growth in the SYK model~\cite{roberts2018operator}.
Taking the continuum limit of the master equation, we get
\bea\label{eq:meq_earlyt}
    \frac{\partial f(h,t)}{\partial t} = \frac{16B}{L}\left(-2\right)\partial_h[hf(h,t)].
\eea
After multiplying Eq.~\eqref{eq:meq_earlyt} by $h$ on both sides and integrating, we see that the mean size obeys
\bea\label{eq:meq_earlyt_2}
    \partial_t\avg{h} = \frac{-32B}{L}\left[
    h^2f(h,t)\Big|_0^L - \avg{h}
    \right]
    \approx \frac{32B}{L}\avg{h}.
\eea
When $B=1/4$ and $h(0)=1$, the solution of Eq.~\eqref{eq:meq_earlyt_2} is
\bea
    \avg{h(t)} = e^{8t/L}.
\eea

\begin{figure}
    \centering
    \includegraphics[width=0.49\columnwidth]{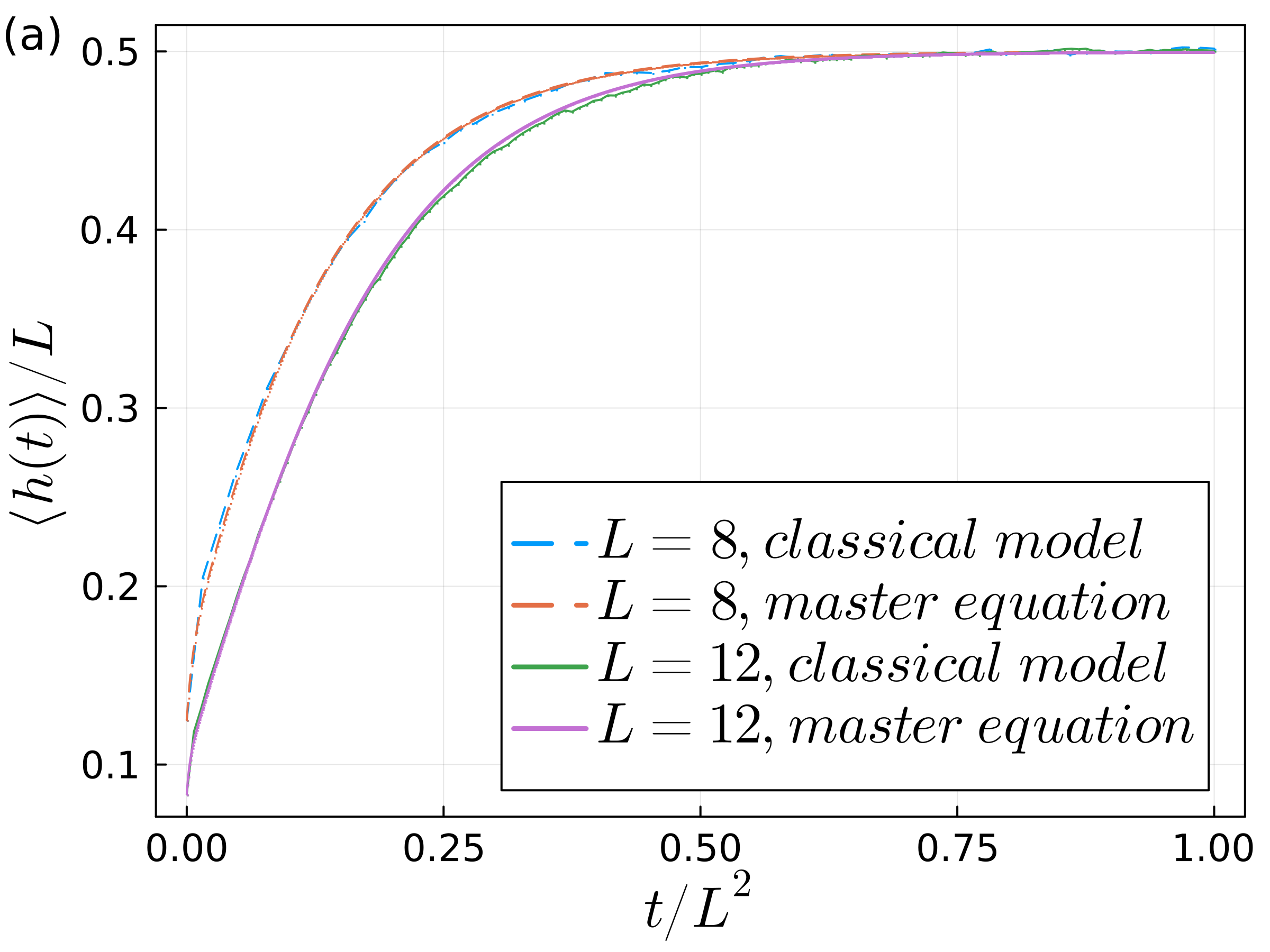}
    \includegraphics[width=0.49\columnwidth]{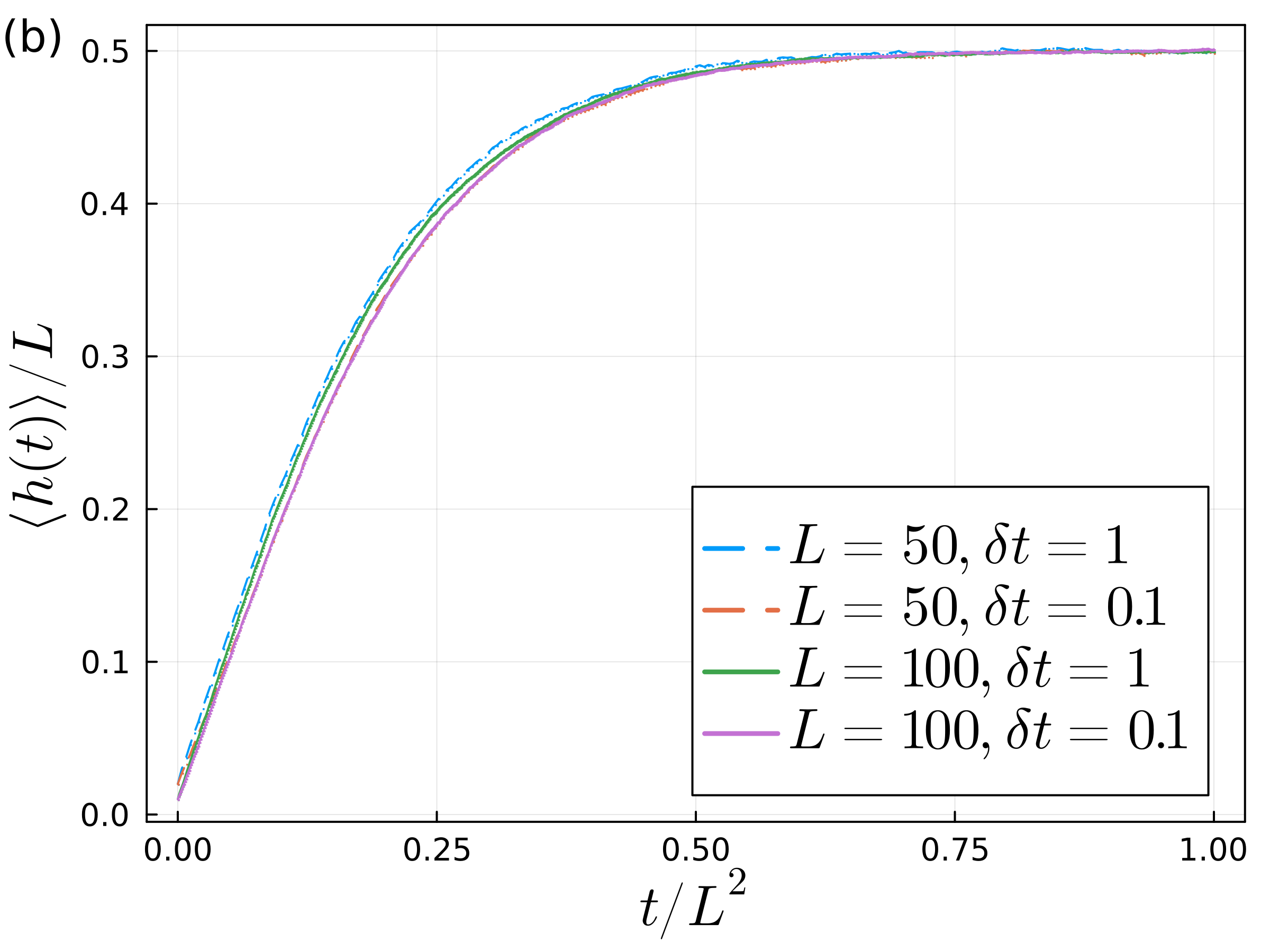}
    \caption{Operator size growth of 1D Brownian circuits with spatial locality. The parameters are the same as those in the main text with $A=B=1/4$.
    (a) Results from simulations of classical stochastic dynamics extracted from the original Brownian circuit with $\delta t=0.5$ (labeled ``classical model'') and results from solving the master equation~\eqref{eq:meqint} (labeled ``master equation'').
    (b) Results from simulations of classical stochastic dynamics extracted from the original Brownian circuit with $\delta t=1$ and $\delta t=0.1$. We observe diffusive dynamics for both cases. In the main text, we use $\delta t=1$.
    }
    \label{add_fig}	
\end{figure}

\begin{figure}
    \centering
    \includegraphics[width=0.5\columnwidth]{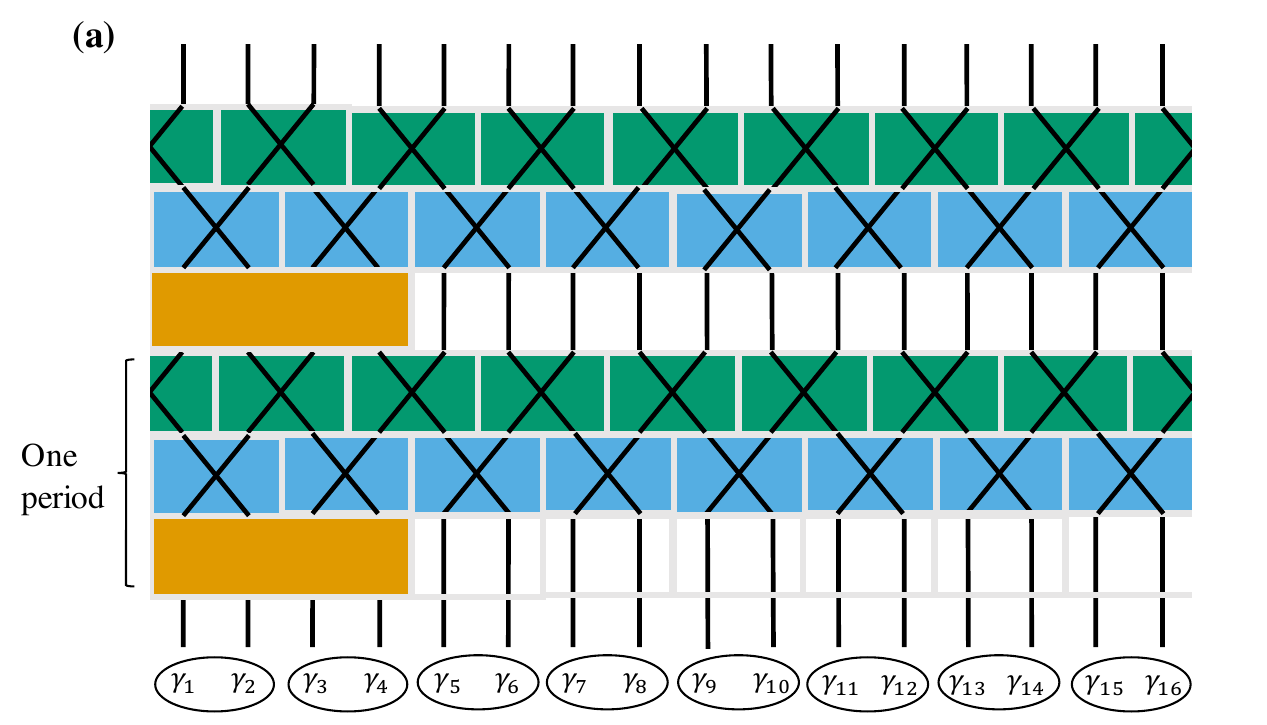}
    \includegraphics[width=0.4\columnwidth]{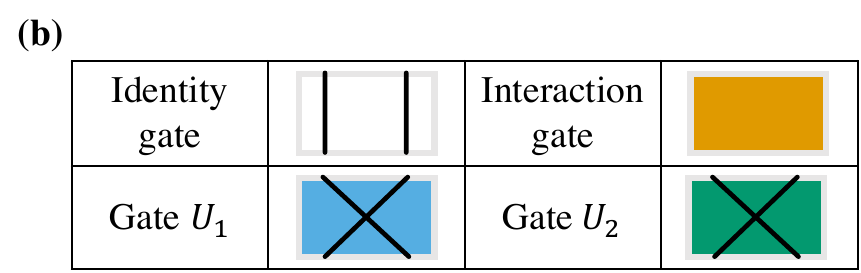}
    \caption{(a) The cartoon picture for the 1D Floquet Clifford circuits in the Majorana fermion representation. The initial state
    is stabilized by pairing up $\{(\gamma_1, \gamma_2),(\gamma_3, \gamma_4), \dots, (\gamma_{2L-1},\gamma_{2L})\}$. In the simulation, we take the periodic boundary condition.
    (b) The possible operations in the Clifford circuits.}
    \label{circuitf}	
\end{figure}

\begin{figure}
    \centering
    \includegraphics[width=0.32\columnwidth]{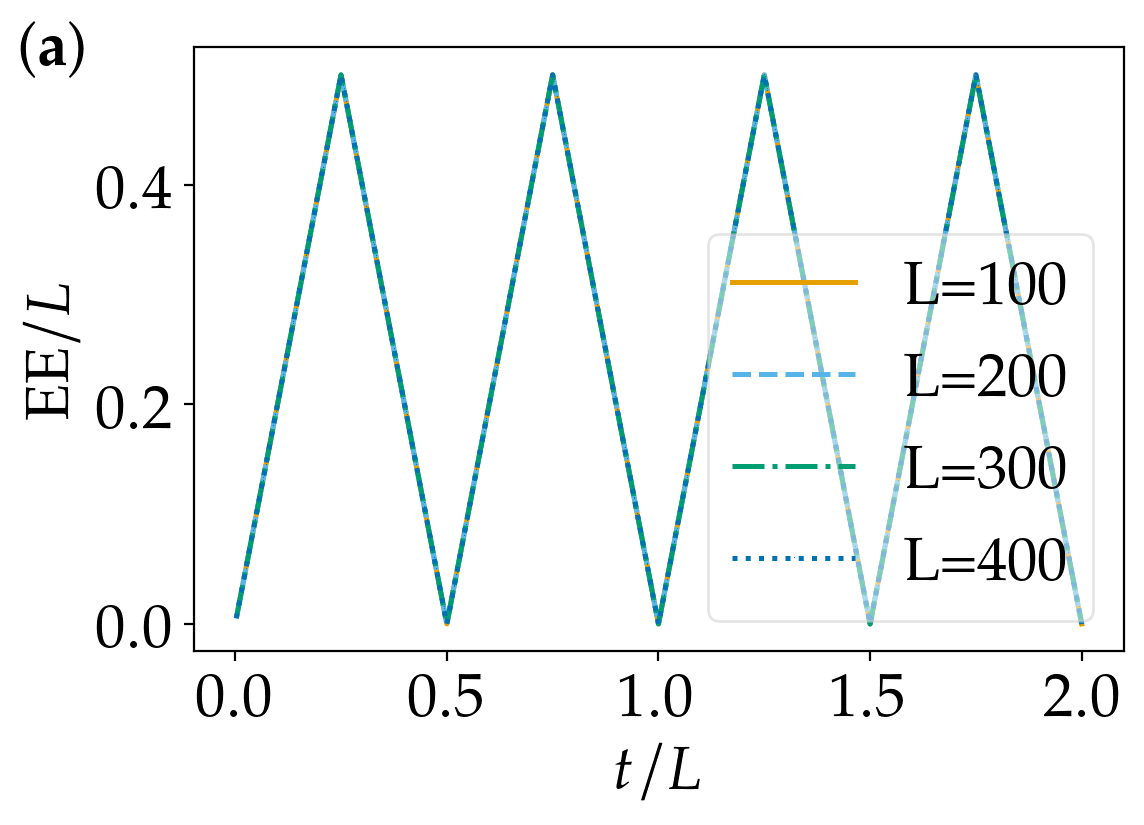}
    \includegraphics[width=0.32\columnwidth]{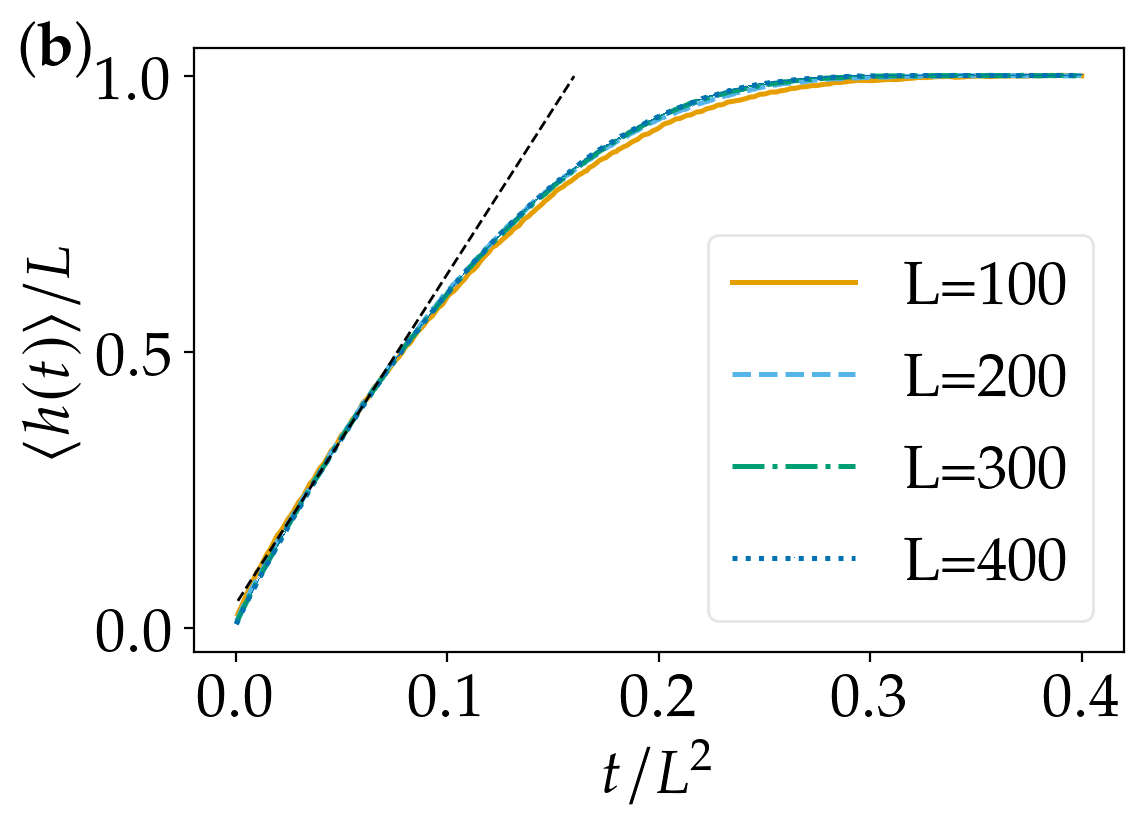}
    \includegraphics[width=0.32\columnwidth]{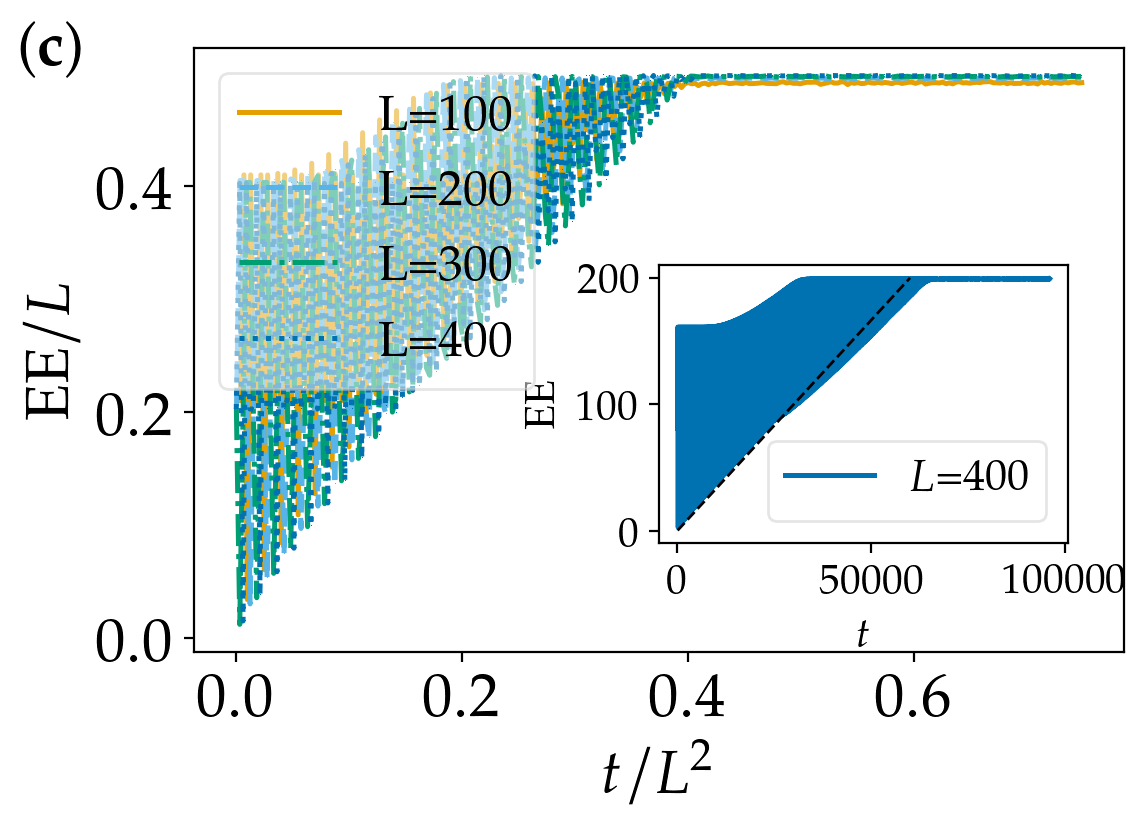}
    \caption{Floquet dynamics in 1D Clifford circuits. 
    Free model: (a) Entanglement growth of subsystem $A$. Free model with a single random two-qubit gate: (b) Operator size growth, (c) Entanglement growth of subsystem $A$. The inset shows the entanglement growth when $L=400$; the black dashed line indicates linear behavior. We bipartite the system into $A$ and $\overline{A}$. The subsystem $A$ is chosen from the $(L/4+2)$-th qubit to the $(3L/4+1)$-th qubit so that the two-qubit interacting gate lies in the middle of the subsystem $\overline{A}$.}
    \label{dataf}
\end{figure}

\section{Comparison between exact results and classical simulations in local Brownian circuits}
For small systems in local Brownian circuits, we find that the classical stochastic dynamics gives the same results as Eq.~\eqref{eq:meqint}.
As shown in Fig.~\ref{add_fig}(a), we find that the classical dynamics' results align with those from the original master equation. However, in such a small system, we cannot observe diffusive dynamics and the data for $L=8$ and $L=12$ do not collapse into a single curve. As we increase the system size, the height distribution $f(\vec{h}, t)$ contains exponentially many components, making a direct numerical study of the master equation impractical. Therefore, we simulate the classical stochastic particle model governed by the master equation. We can observe diffusive dynamics when $L$ is large as shown in Fig.~\ref{add_fig}(b).

\section{Floquet dynamics in 1D Clifford circuits}
We consider a setup with $2L$ Majorana fermions which corresponds to a $L$ qubits system. The initial state is stabilized by $\{i\gamma_1\gamma_2,i\gamma_3\gamma_4, \dots,i\gamma_{2L-1}\gamma_{2L}\}$ and each time period is composed of three steps (also see Fig.~\ref{circuitf}):
\begin{enumerate}
    \item Apply a random two-qubit Clifford gate on the first two qubits, which corresponds to an  interaction term in the Majorana fermion representation.

    \item Perform a swap $U_1 =  e^{i\frac{\pi}{4} (i\gamma_{2i-1}\gamma_{2i}) }$ between Majorana fermion modes $\gamma_{2i-1}$ and $\gamma_{2i}$.

    \item Perform a swap $U_2 =  e^{i\frac{\pi}{4} (i\gamma_{2i}\gamma_{2i+1}) }$ between Majorana fermion modes $\gamma_{2i}$ and $\gamma_{2i+1}$.
\end{enumerate}

For the noninteracting case [evolved without step (1)], the operator size remains the initial value $2$, and the half-system EE oscillates in a linear speed and periodically reaches the maximum possible EE [see Fig.~\ref{dataf}(a)]. The oscillation is a manifestation of the perodic spreading of the two ends of the pair $(\gamma_i, \gamma_j)$.

For the interacting case, the operator size grows nearly linearly at early times and $ \avg{h(t)}/L$ is a scaling function $f(t/L^2)$ for the entire dynamics [see Fig.~\ref{dataf}(b)]. The EE grows in an oscillating way with the mean value growing linearly at early times [see Fig.~\ref{dataf}(c)]. After a timescale of $\sim L^2$, the oscillating behavior of EE disappears, which indicates that information scrambling makes the entanglement structure different from that in the noninteracting case.

\section{Explanation of the coefficient $34/15$ in nonlocal Clifford circuits}
Similar to that in nonlocal Brownian circuits, in the Majorana fermion representation, at early times, when $\avg{h}/L\ll 1$, the probability of only one site among the first four sites being a nonidentity Majorana fermion operator far outweighs the probabilities associated with other nonidentity operators on those initial four sites. The interaction gate can randomly shuffle the $15$ non-identity operators on the initial four sites, with the resulting operator being a size-one operator with probability $4/15$, a size-two operator with probability $6/15$, a size-three operator with probability $4/15$ and a size-four operator with probability $1/15$. Effectively, the interaction gate can increase the size of the size-one operator by a constant $17/15$. Since there are four such size-one operators, we can obtain
\bea
    \frac{d\avg{h}}{dt} = 4\cdot 17/15\cdot \avg{h}/2L = 34/15\cdot \avg{h}/L,
\eea
the solution of which is proportional to an exponential function $\avg{h} \propto e^{34/15\cdot t/L}$.

\end{document}